\documentstyle[12pt]{article}
\def\beq{\begin{equation}}
\def\eeq{\end{equation}}
\def\bea{\begin{eqnarray}}
\def\eea{\end{eqnarray}}
\def\bq{\begin{quote}}
\def\eq{\end{quote}}

\def\PLB{{\it Phys. Lett.} }

\def\NP{{\it Nucl.Phys.} }
\def\PR{{\it Phys.Rev.} }

\def\gappeq{\mathrel{\rlap {\raise.5ex\hbox{$>$}}
{\lower.5ex\hbox{$\sim$}}}}

\def\lappeq{\mathrel{\rlap{\raise.5ex\hbox{$<$}}
{\lower.5ex\hbox{$\sim$}}}}

\begin{document}
\topmargin -0.5cm
\oddsidemargin -0.3cm
\pagestyle{empty}
\begin{center}
{\bf ESTIMATES OF THE HIGHER-ORDER QCD
 CORRECTIONS
 TO $R(s)$, $R_{\tau}$} \\
{\bf AND DEEP INELASTIC SCATTERING SUM RULES}
\footnote{ To be published in Mod. Phys. Lett. A10 (1995) 235.}
\\
\vspace*{1cm}
{\bf Andrei L. Kataev} \footnote{Permanent address: Institute for
Nuclear Research of the Russian Academy of Sciences, 117312 Moscow,
Russia} \\
\vspace{0.3cm}
Theoretical Physics Division, CERN, \\
CH-1211 Geneva 23, Switzerland \\
\vspace{0.5cm}
and \\
\vspace*{0.5cm}
{\bf Valery V. Starshenko} \footnote{Continuation of research
done for a PhD thesis at the Moscow State University, Russia}\\
\vspace*{0.3cm}

Aschara, Germany \\
\vspace*{1cm}
{\bf ABSTRACT} \\
\end{center}
\vspace*{2mm}
\noindent
We present   the attempt to study  the problem  of the
estimates of  higher-order perturbative corrections to physical
quantities in
the Euclidean region. Our considerations
are based on the application of
the scheme-invariant methods, namely the principle of minimal
sensitivity and the effective charges approach. We emphasize, that
in order to obtain the concrete results for the physical quantities
in the Minkowskian region the results of application of this formalism
should be supplemented by the explicite calculations of the effects
of the analytical continuation.
We present the estimates of the order $O(\alpha^{4}_{s})$ QCD
corrections to
the Euclidean quantities: the $e^+e^-$-annihilation $D$-function
and the
deep inelastic scattering sum rules, namely  the
non-polarized and polarized Bjorken sum rules and to the Gross--Llewellyn
Smith sum rule.
 The results for the
$D$-function are further applied to estimate the $O(\alpha_s^4)$
QCD corrections to the Minkowskian quantities
$R(s) = \sigma_{tot} (e^{+}e^{-} \to {\rm hadrons}) /
\sigma (e^{+}e^{-} \to \mu^{+} \mu^{-})$ and
$R_{\tau} = \Gamma (\tau \to \nu_{\tau} + {\rm hadrons}) /
\Gamma (\tau \to \nu_{\tau} \overline{\nu}_{e} e)$.
The problem of the fixation of the uncertainties due to
the $O(\alpha_s^5)$ corrections to the considered quantities
is also discussed.

\vspace*{2cm}



\newpage
\setcounter{page}{1}
\pagestyle{plain}
{\bf I.}~~~
During the last few years, essential progress has been achieved in the
area of the calculation of the next-next-to-leading order ($NNLO$) QCD
corrections to the number of physical quantities.  Indeed, the
complete $NNLO$ $O(\alpha^{3}_{s})$ QCD corrections are known at present
for the characteristics of $e^{+} e^{-} \to \mbox{hadrons}$ process
\cite{aaa},  \cite{bb}, $\tau \to \nu_{\tau} +
\mbox{hadrons}$ decay \cite{cc} $Z^0 \to \mbox{hadrons}$ process
\cite{Z}
  and for the deep inelastic
scattering sum rules, namely the non-polarized Bjorken sum rule (BjnSR)
\cite{ee}, the Gross-Llewellyn Smith sum rule (GLSSR) and the
polarized Bjorken sum rule (BjpSR) \cite{ff}.  Amongst the physical
information provided by these results is the estimate of the
theoretical uncertainties of the corresponding next-to-leading-order
($NLO$) perturbative QCD predictions for
$R(s) = \sigma_{tot}(e^{+} e^{-} \to {\rm hadrons}) /
\sigma (e^{+} e^{-} \to \mu^{+} \mu^{-})$ \cite{ggg},
$\Gamma (Z^{0} \to {\rm hadrons})$ (see. e.g., \cite{hh}),
$R_{\tau} = \Gamma (\tau \to \nu_{\tau} + {\rm hadrons}) /
\Gamma (\tau \to \nu_{\tau} \overline{\nu}_{e} e)$ \cite{cc},
BjnSR \cite{jj}, GLSSR and BjpSR \cite{kk}.

After gaining from
the understanding of the  phenomenological
meaning of the effects of the $NNLO$ corrections to observable
quantities one can of course stop at this level. However, in view
of the fact that the precision of the experimental data for
$R(s)$, $R_{\tau}$, $\Gamma(Z^0\rightarrow hadrons)$ and
 the DIS sum rules is continuously
increasing,
 the questions
about the possible values of the effects of  non-calculated
higher order terms is frequently raised both by theoreticians
and experimentalists. This information can   be useful in
the studies of  several important problems.
It allows to present the concrete  numerical estimate  of the
theoretical uncertainties of the $NNLO$ QCD approximations.
Moreover, any new information,
even empirical,
about the possible values of the higher-order terms might
be useful
in understanding of the general structure of the perturbative
expansions for different quantities
in the concrete renormalization
schemes and  different energy regions.

A certain step in the
direction of the   estimates of the order
$O(\alpha^{4}_{s})$ QCD corrections was made in the case of $R_{\tau}$
in Ref. \cite{cc}.  These estimates are based on the tendency,
observed in Ref. \cite{oo}, of the scheme-dependent uncertainties of
the perturbative QCD predictions for $R(s)$ and $R_{\tau}$ to decrease
as a result of taking into account the order $O(\alpha^{3}_{s})$-terms.
This effect already  occurs at the
$O(\alpha_s^2)$ level. The inclusion of the $O(\alpha_s^3)$
corrections \cite{oo}  confirms this effect and makes it more
vivid.
The foundations of Ref. \cite{oo} were further improved    in the
process of phenomenological studies of the QCD predictions for
$R_{\tau}$, using the explicite separations of the expressions for
the $n$-th order coefficients $r_n^{\tau}$ to the $n$-th order
coefficients $d_n$ of the Euclidean $D$-function and the certain
contributions, which can be explicitely taken into account after
completing the calculations in the $k\leq n-1$ order of the
perturbation theory \cite{pp}.

Notice, that starting from the
$NNLO$ level these contributions contain $\pi^2$-factors which
appear as the result of analytic continuation to the physical
region.
As was demonstarted in Refs.\cite{qq1,qq2},\cite{pp} in order to achieve
better convergence of the corresponding approximants
it is essential to treat them seperately from the coefficients of
the $D$-function.  Even the seperate renormalization-group
\cite{RG} inspired summation of all additional terms in the expression
for $R(s)$ \cite{qq2} and
 $R_{\tau}$ \cite{qq1,pp} was discussed.

This work is the attempt to address the important questions
of the study of the higher-order perturbative approximations
to physical quantities. We will use
 the approach based on the imrovement
formula \cite{rr} of re-expansion of the expressions for the
quantities obtained in the N-th order of perturbation theory
within the principle of minimal sensitivity (PMS) \cite{rr}
and the effective charges (ECH) approach \cite{uu}, which is
equivalent {\it a posteriori} to the scheme-invariant perturbation
theory \cite{vv}.
The main aim of this work, which is based on the considerations
 of Refs. \cite{KatSt1,KatSt2}, is to present the observations
 obtained in the process of the study of the problem of the
possibilities of estimating
 higher-order QCD
 corrections to $R(s)$, $R_{\tau}$ and deep-inelastic scattering
sum rules using the scheme-inariant procedures.

Of course, this approach to the estimation of the values of the
uncalculated higher- order terms cannot be considered as the alternative
to the direct analytical or numerical calculations. However, we hope
that this method can give the concrete prescription for
impression about the possible theoretical
errors due to variation
of the higher order terms.
The first message came from  the fact that the rather bold-guess
application of this method for the estimate of the four-loop corrections
to the expression for $(g-2)_e$ \cite{ss} gave  results, which
turned out to be in  surprisingly good agreement with the latest
results of the direct numerical calculations of Ref. \cite{Kin}
\footnote{A detailed re-consideration of the $(g-2)_{\mu}$
analysis of Ref.
\cite{ss}
and its generalization to the five-loop order will be
presented elsewhere  \cite{fut}.}. The second
argument in favour of this procedure came from its successful
application for the analysis of the Drell-Yan
cross section at the $O(\alpha_s^2)$-level \cite{Aurenche}.
 Moreover, as
will be demonstrated in our work, the re-expansion formalism of
Ref. \cite{rr} is also working quite well in QCD at the $NNLO$
for at least three independent quantities, namely for the
$e^+e^-$-annihilation $D$-function and the non-polarized and polarized
Bjorken sum rules. Therefore, we prefer to consider all these
 facts
 as an arguments in favour of more detailed studies of the
intrinsic features of this approach.

In this work we further develop the re-expansion formalism of Ref.
\cite{rr}, deriving  new terms in the corresponding improvement
formula. The previously-known terms are used to obtain the estimates
of the next-to-next-to-next-to-leading order ($N^3LO$) QCD
 corrections to the $D$-function, the non-polarized
Bjorken sum rule and polarized Bjorken sum rule, which is closely
related by the structure of the corresponding perturbative series
to the Gross--Llewellyn Smith sum rule.
 We will  use the
results obtained for the $D$-function to estimate the effects of the
$N^3LO$ corrections to the perturbative series for the Minkowskian
quantities $R(s)$ and $R_{\tau}$ by adding the explicitly
calculable terms, previously discussed in
Refs.\cite{Bj,pp}, supplementing thus the related
considerations of Refs. \cite{rr}, \cite{ww}
by the additional input information.
The derived
new term in the improvement formula is applied
to touch on the problem of fixing
the values of the $O(\alpha_s^5)$-corrections to the analysed
quantities. We also specify the scheme which is related to our
considerations and emphasize that the structure of the perturbative
expansions in the Euclidean and Minkowskian regions  differ
essentially.

{\bf II.}~~~
Consider first the order $O(a^{N})$ approximation of a Euclidean
physical quantity
\beq
D_{N} = d_{0} a(1 + \sum^{N-1}_{i=1} \, d_{i} a^{i})
\label{1}
\eeq
with $a = \alpha_{s}/\pi$ being the solution of the corresponding
renormalization group equation for the $\beta$-function which is
defined as
\beq
\mu^{2} \frac{\partial a}{\partial \mu^{2}} =
\beta (a) = - \beta_{0} a^{2}
(1 +  \sum^{N-1}_{i=1} \, c_{i} a^{i})\ .
\label{2}
\eeq
In QCD,
in the process of the concrete calculations of the coefficients
$d_{i}, i \geq 1$ and $c_{i}, i \geq 2$, the $\overline{MS}$ scheme is
commonly used.  However, this scheme is not the unique prescription for
fixing the RS ambiguities (for the recent discussions see e.g., Ref.
\cite{ww}).

The PMS \cite{rr} and ECH \cite{uu} prescriptions stand out from
various methods of treating scheme-dependence ambiguities.
 Indeed, they are based on the conceptions of the
scheme-invariant quantities, which are defined as the combinations of
the scheme-dependent coefficients in Eqs. (1) and (2).  Both these
methods pretend to be
  ``optimal" prescriptions, in the sense that they might
 provide better convergence of the corresponding approximations in
the non-asymptotic regime, and thus allow an estimation of the
uncertainties of the perturbative series
in the definite order of perturbation theory.  Therefore,
applying these
 ``optimal" methods,  one can try to estimate the
effects of the order $O(a^{N+1})$-corrections starting from the
approximations $D^{opt}_{N} (a_{opt})$ calculated in a certain
``optimal" approach \cite{rr}, \cite{ss}, \cite{tt}.
This idea is closely
related to the QED technique of Ref. \cite{zz}, which was used to
predict  the renormalization-group controllable $ln(m_{\mu}/m_e)$-terms
in the series for
$(g-2)_{\mu}$ from the expression of $(g-2)_{\mu}$ through the
effective coupling constant $\bar{\alpha} (m_{\mu} / m_{e})$.
In our work
work we are using this technique to estimate also
the constant terms of  the higher-order corrections in QCD.

Let us following the considerations of Ref. \cite{rr}
re-expand $D_{N}^{opt} (a_{opt})$ in terms of the coupling
constant $a$ of the particular scheme
\beq
D_{N}^{opt} (a_{opt}) = D_{N} (a) + \delta D_{N}^{opt} a^{N+1}
\label{3}
\eeq
where
\beq
\delta D_{N}^{opt} = \Omega_{N}(d_{i}, c_{i}) -
\Omega_{N} (d_{i}^{opt}, c_{i}^{opt})
\label{4}
\eeq
are the  numbers which simulate the coefficients of the
order $O(a^{N+1})$-corrections to the physical quantity, calculated in
the particular initial scheme, say the $\overline{MS}$-scheme.  The
coefficients $\Omega_{N}$ can be obtained from the following system of
equations:
\bea
\frac{\partial}{\partial \tau}
(D_{N} + \Omega_{N} a^{N+1}) =
O(a^{N+2}), \nonumber \\
\frac{\partial}{\partial c_{i}}
(D_{N} + \Omega_{N} a^{N+1}) =
O(a^{N+2}),\ i \geq 2
\label{5}
\eea
where the parameter $\tau = \beta_{0} \ell n (\mu^2 / \Lambda^2)$
represents the freedom in the choice of the renormalization point
$\mu$. The conventional scale parameter $\Lambda$ will not explicitly
appear in all our final formulas.

The explicit form of the coefficients $\Omega_{2}$ and $\Omega_{3}$ in
which we will be interested can be obtained by the solution of the
system of equations (5), following the lines of Ref. \cite{rr}.  We
present here  the final already known expressions \cite{rr}:
\beq
\Omega_{2} = d_{0}d_{1} (c_{1} + d_{1}),
\label{6}
\eeq
\beq
\Omega_{3} = d_{0}d_{1} (c_{2} - \frac{1}{2} c_{1}d_{1}
-2d_{1}^{2} + 3d_{2})\ .
\label{7}
\eeq
and the new term $\Omega_{4}$ evaluated by us:
\beq
\Omega_{4}=\frac{d_0}{3} ( 3c_{3}d_1+c_2d_2-4c_2d_1^2+2c_1d_1d_2
-c_1d_3+14d_1^4-28d_1^2d_2+5d_2^2+12d_1d_3 )
\label{new1}
\eeq
which reproduces the renormalization-group controllable logarithmic
terms at the five-loop level \cite{Kat1}.

It should be stressed that in the ECH approach $d_{i}^{ECH} \equiv 0$
for all $i \geq 2$.  Therefore one gets the following expressions for
the $NNLO$ and $N^3LO$ corrections in Eq. (3):
\beq
\delta D_{2}^{ECH} = \Omega_{2} (d_{1}, c_{1})
\label{8}
\eeq
\beq
\delta D_{3}^{ECH} = \Omega_{3}(d_{1}, d_{2}, c_{1}, c_{2})
\label{9}
\eeq
\beq
\delta D_{4}^{ECH}=\Omega_4(d_1,d_2,d_3,c_1,c_2,c_3)
\label{new2}
\eeq
where $\Omega_{2}$, $\Omega_{3}$ and $\Omega_{4}$
are defined in Eqs. (6),(7) and (8) respectively.

It is worth emphasizing that the general expressions
for the correction coefficients $\delta D_N=
d_0d_{N}$ in Eqs. (9), (10), (11)
can be obtained from the following exact relation
for the process dependent but scheme-independent quantities
\beq
\frac{c_{N}^{ECH}}{N-1}=\frac{c_{N}}{N-1}+d_{N}-\frac{\Omega_{N}}{d_0}
\label{schinv}
\eeq
which are related to the coefficients $c_N^{ECH}$
of the ECH $\beta$-functions defined as $\beta_{eff}(a_{ECH})=
-\beta_0a_{ECH}^2(1+c_1a_{ECH}+\sum_{i \geq 2}c_i^{ECH}a_{ECH}^i)$.
Therefore, the expressions for the corrections $\delta D_N^{ECH}$
are the \underline{exact} numbers  which are related to the
special scheme. This scheme
is identical to the $\overline{MS}$ scheme
at the lower order levels and is defined by the condition
$c_{N}=c_{N}^{ECH}$ at the N-th order, where $c_N^{ECH}$ are
considered as unknown numbers.

In order to find similar corrections to Eq.(3) in the N-th order of
perturbation theory starting from the PMS approach \cite{rr}, it is
necessary to use the relations obtained in Ref. \cite{yy} between the
coefficients $d_{i}^{PMS}$ and $c_{i}^{PMS} \; (i \geq 1)$ in the
expression for the order $O(a^{N}_{PMS})$ approximation
$D^{PMS}_{N} (a_{PMS})$ of the physical quantity under consideration.
The corresponding corrections have the following form:
\beq
\delta D_{2}^{PMS} = \delta D_{2}^{ECH} + \frac{d_{0} c^{2}_{1}}{4}
\label{10}
\eeq
\beq
\delta D_{3}^{PMS} = \delta D_{3}^{ECH}\ .
\label{11}
\eeq
Notice the identical coincidence of the $N^3LO$ corrections obtained
starting from both the PMS and ECH approaches.  A similar observation
was made in Ref. \cite{ss} using different (but related) considerations.

In the fourth order of  perturbation theory the additional
contribution to $\delta D_{4}^{PMS}$ has more complicated structure.
Indeed, the expression for $\Omega_4(d_i^{PMS},c_i^{PMS})$ in
Eq. (4) reads:
\beq
\Omega_4(d_i^{PMS},c_i^{PMS})=\frac{d_0}{3} [ \frac{1}{4}c_1c_3^{PMS}
-\frac{4}{81}(c_2^{PMS})^2-\frac{5}{81}c_1^2c_2^{PMS}+\frac{7}{648}
c_1^4]
\label{omega4}
\eeq
where
\beq
c_2^{PMS}=\frac{9}{8}(d_2+c_2-d_1^2-c_1d_1+\frac{7}{36}c_1^2)
+O(a)
\label{c2pms}
\eeq
and
\beq
c_3^{PMS}=4(d_3+\frac{1}{2}c_3-c_2d_1-3d_1d_2+2d_1^3)
+\frac{1}{2}c_1(d_2+c_2+3d_1^2-c_1d_1+\frac{1}{108}c_1^2)+O(a)
\label{c3pms}.
\eeq
The expressions for Eqs. (15)- (17) are  pure numbers, which
do not depend on the choice of the initial scheme.
Note that we have checked that in the case of the consideration
of the perturbative series for  $(g-2)_{\mu}$
the numerical values of $\Omega_4(d_i^{PMS},c_i^{PMS})$ are
small and thus the {\it a posteriori} approximate equivalence
of the ECH and PMS approaches is preserved for the quantities
under consideration at this level also \cite{fut}. We think that
this feature is also true in QCD.

{\bf III.}~~~
Consider now   the familiar characteristic of the $e^{+}e^{-} \to
\gamma \to {\rm hadrons}$ process, namely the $D$-function defined in
the Euclidean region:
\beq
D(Q^{2}) = Q^{2} \int^{\infty}_{0} \, \frac{R(s)}{(s+Q^{2})^{2}} \, ds
\label{12}
\eeq
Its perturbative expansion has the following form:
\bea
D(Q^{2})  =  3 \Sigma Q^{2}_{f} [1 + a + \sum _{i\geq 1} \,
 d_{i}a^{i+1}  ]
 +  (\Sigma Q_{f})^{2} [ \tilde{d}_{2} a^{3} + O(a^{4}) ]
\label{13}
\eea
where $Q_{f}$ are the quark charges, and the structure proportional to
$(\Sigma Q_{f})^{2}$ comes from the light-by-light-type diagrams. The
coefficients $d_{1}$ and $d_{2}, \tilde{d}_{2}$ were calculated in the
$\overline{MS}$-scheme in Refs. \cite{ggg} and \cite{aaa,bb}
 respectively.
They have the following numerical form:
\bea
d_{1}^{\overline{MS}} & \approx & 1.986 - 0.115 f \nonumber \\
d_{2}^{\overline{MS}} & \approx & 18.244 - 4.216 f + 0.086 f^{2} ,\
\tilde{d}_{2} \approx -1.240\ .
\label{14}
\eea

Following the proposals of Ref. \cite{aai}, we will treat the
light-by-light-type term in Eq. (19) separately from the ``main"
structure of the $D$-function, which is proportional to the
quark-parton expression $D^{QP}(Q^{2}) = 3 \Sigma Q^{2}_{f}$. In fact,
one can hardly expect that it is possible to ``predict'' higher-order
coefficients $\tilde{d}_{i}, i \geq 3$ of the second structure in Eq.
(19) using the only explicitly-known term $\tilde{d}_{2}$.  Therefore
we will neglect the light-by-light-type structure as a whole in all our
further considerations.  This approximation is supported by the
relatively tiny contribution of the second structure of Eq. (19) to the
final $NNLO$ correction to the $D$-function.

The next important ingredient of our analysis is the QCD
$\beta$-function (2), which is known in the MS-like schemes at the $NNLO$
level \cite{bbi}.  Its corresponding coefficients read
\bea
\beta_{0} & = & (11 - \frac{2}{3} f) \frac{1}{4} \approx
2.75 - 0.167 f \nonumber \\
c_{1} & = & \frac{153 - 19f}{66 - 4 f} \nonumber \\
c_{2}^{\overline{MS}} & = & \frac{77139 - 15099f + 325 f^{2}}
{9504 - 576 f}\ .
\label{16}
\eea
Using now the perturbative expression for the $D$-function, one can
obtain the perturbtive expression for $R(s)$, namely
\bea
R(s)  =  3 \Sigma Q^{2}_{f} [1 + a_{s} + \sum_{i\geq 1} \,
r_{i} a^{i+1}_{s} ]
 +  (\Sigma Q_{f})^{2} [ \tilde{r}_{2} a^{3}_{s} + ... ]
\label{17}
\eea
where $a_{s} = \bar{\alpha}_{s} / \pi$, and
\bea
r_{1} & = & d_{1} \nonumber \\
r_{2} & = & d_{2} - \frac{\pi^{2} \beta^{2}_{0}}{3}, \,\
\tilde{r}_{2} = \tilde{d}_{2} \nonumber \\
r_{3} & = & d_{3} - \pi^{2} \beta^{2}_{0}
(d_{1} + \frac{5}{6} c_{1})\ .
\label{18}
\eea
The corresponding $\pi^{2}$-terms come from the analytic continuation
of the Euclidean result for the $D$-function to the physical region.
The effects of the higher-order $\pi^{2}$-terms were discussed in
detail in Ref. \cite{Bj}. For example, the corresponding espression for
the $r_4$-term reads :
\bea
r_{4} &=& d_{4} -\pi^2\beta_0^2 (2d_2+\frac{7}{3}c_1d_1
+\frac{1}{2}c_1^2+c_2)+\frac{\pi^4}{5}\beta_0^4
\label{r4}
\eea

The perturbative expression for $R_{\tau}$ is defined as
\bea
R_{\tau}  =  2 \int_{0}^{M^{2}_{\tau}} \, \frac{ds}{M^{2}_{\tau}} \,
(1 - s/M^{2}_{\tau})^{2} \, (1 + 2s/M^{2}_{\tau}) \tilde{R}(s)
 \simeq  3[1 + a_{\tau} + \sum_{i\geq 1} \,
r_{i}^{\tau} a^{i+1}_{\tau} ]
\label{19}
\eea
where $a_{\tau} = \alpha_{s} (M^{2}_{\tau}) / \pi$ and $\tilde{R} (s)$
is $R(s)$ with
with $f = 3, (\Sigma Q_{f})^{2} = 0, 3 \Sigma Q^{2}_{f}$ substituted
for $3 \Sigma \mid V_{ff'} \mid^{2}$ and $\mid V_{ud} \mid^{2} +
\mid V_{us} \mid^{2} \approx 1$.

It was shown in Ref. \cite{pp} that it is convenient to
express the coefficients of the
series (25) through  those ones of the series (19) for the
$D$-function in the following form :
\bea
r^{\tau}_{1} & = &
d_{1}^{\overline{MS}} (f = 3) +  g_1 (f=3)  \nonumber \\
r^{\tau}_{2} & = &
d_{2}^{\overline{MS}} (f = 3) +  g_2 (f=3) \nonumber
\\
r^{\tau}_{3} & = &
d_{3}^{\overline{MS}} (f = 3) + g_3 (f=3)
\label{20}
\eea
where in our notations
\bea
g_1 & = & -\beta_0I_1 = (f=3)=3.563 \nonumber \\
g_2 & = & -[2d_1+c_1]\beta_0 I_1+\beta_0^2 I_2=(f=3)=19.99 \nonumber \\
g_3 & = & -[3d_2+2d_1c_1+c_2]\beta_0 I_1+
[3d_1+\frac{5}{2}c_1]\beta_0^2 I_2
-\beta_0^3 I_3=(f=3)=78.00
\label{gn}
\eea
and $I_{k}$ are defined and calculated
in Ref. \cite{pp}. Their  analytical expressions
read: $I_1=-19/12$, $I_2=265/72-\pi^2/3$ and $I_3=-3355/288+
19\pi^2/12$. One of the pleasant features of Eqs.(\ref{gn}) is that
they are absorbing all effects of the analytical continuation.

Following the lines of Ref.\cite{pp} we derive the corresponding
expression for the coefficient $r^{\tau}_{4}$:
\bea
r^{\tau}_{4} = d_4^{\overline{MS}}(f=3) + g_4 (f=3)
\label{r4tau}
\eea
where
\begin{eqnarray}
g_4&=&-[4d_3+3d_2c_1+2d_1c_2+c_3]\beta_0I_1+[6d_2+7c_1d_1+
\frac{3}{2}c_1^2+3c_2]\beta_0^2I_2   \nonumber \\
&& -[4d_1+\frac{13}{3}c_1]\beta_0^3I_3
+\beta_0^4I_4 = (f=3) \nonumber \\
&&=3.562c_3(f=3)+14.247d_3(f=3)-466.73
\label{g4}
\end{eqnarray}
and $I_4=41041/864-265\pi^2/36+\pi^4/5 \approx -5.668$.

In order to estimate the values of the order $O(a^{3})$, $O(a^{4})$
and $O(a^5)$
corrections to $R(s)$ and $R_{\tau}$, we will apply Eqs. (6) - (11) in
the Euclidean region to the perturbative series for the $D$-function
and then obtain the  estimates we are interested in
using Eqs. (23), (24), (26)-(29).
This is the new ingredient of this analysis, which distinguishes
it from the related ones of Refs. \cite{rr}, \cite{ww}.

We now recall the perturbative expression for the non-polarized Bjorken
deep-inelastic scattering sum rule
\bea
BjnSR  = \int^{1}_{0} \, F_{1}^{\bar{v}p - vp} (x, Q^{2}) dx
 =  1 - \frac{2}{3} a (1 + \sum_{i\geq 1} \,
 d_{i}a^i )
\label{21}
\eea
where the coefficients $d_{1}$ and $d_{2}$ are known in the
$\overline{MS}$
scheme from the results of calculations of Ref. \cite{jj} and Ref.
\cite{ee} respectively:
\beq
d_{1}^{\overline{MS}} \approx 5.75 - 0.444 f
\label{22}
\eeq
\beq
d_{2}^{\overline{MS}} \approx 54.232 - 9.497 f + 0.239 f^{2}\ .
\label{23}
\eeq
The expression for the polarized Bjorken sum rule BjpSR has the
following form:
\bea
BjpSR  = \int^{1}_{0} \, g_{1}^{ep-en} (x, Q^{2}) dx
 =  \frac{1}{3} \mid \frac{g_A}{g_V} \mid \,
[ 1-a (1 + \sum_{i\geq 1} \, d_{i} a^{i}) ]
\label{24}
\eea
where the coefficients $d_{1}$ and $d_{2}$ were explicitly calculated
in the $\overline{MS}$ scheme in
Refs. \cite{kk} and \cite{ff} respectively.
The results of these calculations read
\beq
d_{1}^{\overline{MS}} \approx 4.583 - 0.333 f
\label{25}
\eeq
\beq
d_{2}^{\overline{MS}} \approx 41.440 - 7.607 f + 0.177 f^{2}\ .
\label{26}
\eeq
It should be stressed that since deep inelastic scattering sum rules
are defined in the Euclidean region, we can directly apply to them the
methods discussed in Section 2 without any additional modifications.

It is also worth emphasizing that, in spite of the identical
coincidence of the $NLO$ correction to the Gross-Llewellyn Smith sum rule
$GLSSR  =  (1/2)\int^{1}_{0} F_3^{\overline{\nu}p+\nu p}
(x,Q^2)dx$
with the result of Eq. (34) \cite{kk}, the corresponding $NNLO$
 correction
differs from the result of Eq. (35) by the contributions of the
light-by-light-type terms typical of the GLSSR \cite{ff}.
Since these light-by-light-type
terms appear for the first time at the $NNLO$, it is impossible to
predict the value of the light-by-light-type contribution at the $N^3LO$
level using the corresponding $NNLO$ terms as the input information.
However, noticing that at the $NNLO$ level
the corresponding light-by-light-type contributions are small \cite{ff},
we will assume that the similar contributions
are small at the $N^3LO$ level also. Only after this assumption can
our estimates of the $NNLO$ and $N^3LO$ corrections to the BjpSR  be
considered also as the estimates of the corresponding
corrections in the perturbative series for the GLSSR . Note, that
 the
Pad\'e predictions of the order $O(a^{4})$ contributions
to the GLSSR \cite{ddi}, which do not take into account the necessity
of the careful considerations of the light-by-light-type terms,
deserve more detailed considerations. To our point of view,
 it is better to neglect the small light-by-light
contribution as the whole and to consider the problem of estimates
of the higher order corrections to BjpSR and GLSSR simultaniously,
but not seperately, as was done in Ref. \cite{ddi}.

{\bf IV.}~~~
The estimates of the coefficients of the order $O(a^{3})$ and
$O(a^{4})$ QCD corrections to the $D$-function, $R(s)$, BjnSR and
BjpSR/GLSSR, obtained following
the discussions of Section 2 with the help of
the results summarized in Section 3, are presented in Tables 1 - 4
respectively.  Due to the complicated $f$-dependence of the
coefficients $\Omega_{2}, \Omega_{3}$ in Eqs. (6) and (7), we are
unable to predict the explicit $f$-dependence of the corresponding
coefficients in the form respected by perturbation theory.  The
results are presented for the fixed number of quark flavours $1 \leq f
\leq 6$.
The estimates of
the $NNLO$ corrections, obtained starting from both the ECH and PMS
approaches, are in qualitative agreement with the results of
 the explicit calculations. The best agreement is achieved for
$f=3$ numbers of flavours.

Using the results of Table 1 and Eqs. (25),(26)
 we get the following estimates of the $NNLO$ coefficients
 $R_{\tau}$:
$ ( r^{\tau}_{2}  )^{est}_{ECH} \approx 25.6$
and $\left ( r^{\tau}_{2} \right )^{est}_{PMS} \approx 24.8$ .
One can see the agreement with the explicitly calculated result
$ ( r^{\tau}_{2}  )_{\overline{MS}} = 26.366$ .
Considering this agreement as the additional {\it a posteriori} support
of the methods used, we use the estimate of the $N^3LO$ coefficient for
the $D$-function with $f = 3$ numbers of flavours as presented in Table
1, namely:
\beq
d^{est}_{3}(f=3) = 27.5
\label{30}
\eeq
and estimate the value of the $N^3LO$ coefficient of $R_{\tau}$:
$\left (r^{\tau}_{3} \right )^{est} \approx 105.5$ .

As follows from Eq. (12) the results given in Tables 1-4 correspond
to the ECH-inspired variant of the $\overline{MS}$ scheme.
However, we hope that there is some meaning in the relation
$c_N=c_N^{ECH}$ used to obtain these numbers. Indeed, it is known
from the explicite QED calculations of Refs. \cite{beta} that
the difference between the numerical values of the higher order
coefficients of the QED $\beta$-function in different schemes is
decreasing at the four-loop level. So, we do not exclude the
situation that the difference between scheme-invariants $c_N^{ECH}$
of different QCD quantities, which probably have the factorial
growth $c_N^{ECH} \sim AN!$, will also decrease in the higher orders
of perturbation theory, and that it might happen that $c_N\approx
c_N^{ECH}$.
 If so, the estimates of Tables 1-4
 reveal the structure of the perturbative series for
physical quantities in the  fixed scheme.

In order to address the problem of fixing the values of the
$O(a^5)$ QCD corrections to the considered Euclidean quantities
we apply Eqs. (8), (11) with the explicitely calculated $NLO$
and $NNLO$ coefficients $d_1,d_2,c_1,c_2$ in the $\overline{MS}$
scheme and use the determined
from Eqs. (7), (10) estimates of the $N^3LO$ coefficients $d_3$.
In order to obtain the estimates of the next-to-next-to-next-to-next-
to-leading order ($N^4LO$) coefficients of $R(s)$ and $R_{\tau}$,
which are related to the $N^4LO$ coefficient $d_4$ of the $D$-function,
the explicitly calculated terms in Eq.(24) and Eq. (29) are taken
into account.

However, the expressions for $\Omega_4$ in Eq. (8) and $g_4$
in Eq. (29) depend also on the  four-loop
coefficient $c_3$ of the QCD $\beta$-function, which is unknown
at present. Therefore, in
Tables 1-4 we present the estimates
 of the combinations
$d_4-d_1c_3$ and $r_4-r_1c_3$. The existing uncertainty in the
value of $N^3LO$ coefficients $d_3$ is fixed by the assumption
that the real values of these coefficients do not significantly
differ from the $N^3LO$ estimates, obtained by us. However, in
the case  of the
$D$-function with $f=3$ numbers of flavours we present also the
more detailed expression of $d_4$, which follows from Eqs.(11),(8):
\beq
(d_4)_{ECH}^{est}(f=3)= 1.64 c_3 (f=3) + 5.97 d_3(f=3)-52.8  .
\label{de}
\eeq
Taking into account the negative contributions into
 the coefficient $g_4$
in Eq. (28) we get the following estimate of the $N^4LO$ coefficient
of $R_{\tau}$ :
\beq
(r_4^{\tau})_{ECH}^{est}=5.2 c_3 (f=3) +20.22 d_3(f=3) - 519.5 \ .
\label{rtau}
\eeq
Of course, the best way of fixing the value of $c_3$ would be its
explicite calculation. However, at the current level of art one is
free to invent his own way of fixing this part of the ambiguities
of the less substantiated than $N^3LO$ estimates of the $N^4LO$
terms \footnote{Another kind of the uncertainties is related to
the possible deviations of the real values of the $N^3LO$ corrections
from the presented by us estimates.}. We will use here the bold guess
estimate $c_{3}(f=3) \approx c_{2}(f=3)^2/c_{1}(f=3)\approx 11$
, which is motivated by the good agreement
of our results of Eq. (36) with the  approximation
$d_3\approx d_2^2/d_1\approx 25$,
 previously
used in Ref. \cite{pp} to fix the value of the $N^3LO$ coefficient of the
$D$-function for $f=3$
. Combining it with the estimate of Eq. (36) we get the
following  estimate of the $N^4LO$
coefficients of $R_{\tau}$:
$(r_4^{\tau})^{est} \approx 93$.

{\bf V.}~~~
We are now ready to discuss the main otcomes of our analysis.

1.~~
The estimates   of the $NNLO$ corrections to the $D$-function, BjnSR and
BjpSR/GLSSR are in qualitative agreement
with the results of the explicit
calculations of Refs. \cite{aaa,bb}, \cite{ee} and \cite{ff} and respect
the tendency of the corresponding coefficients to decrease with
increasing number of flavours.

2.~~
The best agreement of the $NNLO$ estimates with the exact results is
obtained for the case $f = 3$.  This fact
supports the application of the method used for estimating the $NNLO$ and
$N^3LO$ corrections to $R_{\tau}$.

3.~~
 Notice that since the methods used correctly
reproduce the renormalization-group-controllable terms \cite{zz},
\cite{rr}, the transformation of the coefficients $d_1$, $d_2$
from the $\overline{MS}$ scheme to other
variants of the MS-like scheme will not spoil the qualitative agreement
with the results of the explicit calculations.

4.~~
The comparison of the results of Table 1  with those
 of Table 2 demonstrate that the $\pi^{2}$ effects give
dominating contributions to the coefficients of $R(s)$.

5.~~ Amongst other outcomes
are the estimates of the $N^3LO$ and $N^4LO$
corrections to $R_{\tau}$
 and $R(s)$  for $f = 5$ numbers of flavours.  Taking $\alpha_{s}
(M_{Z}) \approx 0.12$, we get the estimate of the corresponding $N^3LO$
contribution to both $\Gamma (Z^{0} \to {\rm hadrons})$ and
$\Gamma (Z^{0} \to \bar{q}q)$:
\beq
(\delta \Gamma_{Z^{0}})_{N^3LO}
\approx -97 (a(M_{Z}))^{4} \approx -2 \times
10^{-4}\ .
\label{32}
\eeq
It is of the order of magnitude of other corrections included in the
current analysis of LEP data (see, e.g., \cite{eei}).
Using the estimates presented in Table 2 and assuming that
$c_{3}(f=5)=c_{2}(f=5)^2/c_{1}(f=5)\approx 1.715$ one can also estimate
 the corresponding $N^4LO$ contributions, namely
\beq
(\delta \Gamma_{Z^{0}})_{N^4LO}
\approx 70 (a(M_{Z}))^{5} \approx 5 \times
10^{-6}
\label{znew}
\eeq
The   way of fixation of the value of $c_3$ used above
is not applicable for the case of $f=6$ since we expect
that in this case its real value is negative.

6.~~
Taking $\alpha_{s}(M_{\tau}) \approx 0.36$ \cite{cc}, we get the
numerical estimate of the $N^3LO$ contribution to $R_{\tau}$:
\beq
(\delta R_{\tau})_{N^3LO} \approx 105.5 \, a^{4}_{\tau}
\approx 1.8 \times 10^{-2}\ .
\label{33}
\eeq
It is larger than the recently calculated power-suppressed perturbative
\cite{ffi} and non-perturbative \cite{ggi}
contributions to $R_{\tau}$.
The possible contributions of the $N^4LO$ corrections are
significantly smaller:
\beq
(\delta R_{\tau})_{N^4LO} \approx 93  a^{5}_{\tau}
\approx 1.8 \times 10^{-3}\ .
\label{33}
\eeq

7.~~
Notice, that in spite of the possible $N!$ growth of the
coefficients of the $D$-function, the additianal $\pi^2$-dependent
contributions are shadowing down this possible growth in the
Minkowskian region. In order to consider this effect in higher
order levels we applied the RG-technique and
derived the corresponding relations between
the order $O(a^6)$ coefficients of the $D$-function and
$R(s)$ and $R_{\tau}$. In the case of $R(s)$ the relation reads
\begin{eqnarray}
r_5&=&d_5-\frac{\pi^2}{3}\beta_0^2\big(10d_3+\frac{27}{2}c_1d_2
+4c_1^2d_1+\frac{7}{2}c_1c_2+8c_2d_1+\frac{7}{2}c_3\big)
\nonumber \\
&&+\frac{\pi^4}{5}\beta_0^4\big(5d_1+\frac{77}{12}c_1).
\end{eqnarray}
In the case of $R_{\tau}$ the correction term to the
relation $r_5^{\tau}=d_5(f=3)+g_5(f=3)$ is more complicated
\begin{eqnarray}
g_5&=&-[5d_4+4c_1d_3+3c_2d_2+2c_3d_1+c_4]\beta_0I_1
\nonumber \\
&&+[10d_3+\frac{27}{2}c_1d_2+4c_1^2d_1+8c_2d_1+\frac{7}{2}c_1c_2
+\frac{7}{2}c_3]\beta_0^2I_2 \nonumber \\
&&-[10d_2+\frac{47}{3}c_1d_1+\frac{35}{6}c_1^2+6c_2]\beta_0^3I_3
+[5d_1+\frac{77}{12}c_1]\beta_0^4I_4-\beta_0^5I_5
\end{eqnarray}
where $I_5=-2479295/10368+16775\pi^2/432-19\pi^4/12 \approx
-10.113$. Keeping in the expression for $g_5(f=3)$
all explicitely unknown coefficients we get
\begin{equation}
g_5(f=3)=17.8d_4(f=3)+45.07d_3(f=3)+3.56c_4(f=3)+18.6c_3(f=3)
-8455 .
\end{equation}
Notice the appearence in the expression for  $g_5(f=3)$ of the
huge negative coefficient. The derived expressions
can  mean that the asymptotic structure
of the perturbative expansion in the Minkowskian region differs
essentially from the one in the Euclidean region.

8.~~
The qualitative agreement of the results of Tables 3 and 4 for BjnSR
and BjpSR with the corresponding Pad\'e estimates of Ref. \cite{ddi}
can be considered as the argument in favour of the applicability of
both theoretical methods in the Euclidean region for the concrete
physical applications. Let us stress again that in the process of
these applications the light-by-light-type structures, contributing
 to  the GLSSR, should be treated seperately.

9.~~
Our estimate for $d_{3} (f = 3)$ of Eq. (36) is more definite than the
one presented in Ref. \cite{sli}, namely $d_{3} (f = 3) =
55^{+60}_{-24}$, and than the bold guess estimate $d_{3} (f = 3) = \pm
25$, given in Ref. \cite{pp}.
 The result is in good agreement with the
``geometric progression'' assumption of Ref. \cite{pp}.
 The related estimate of the
$N^3LO$ contribution to $R_{\tau}$ (see Eq. (41)) is more precise than
those presented in Refs. \cite{cc} and \cite{pp}, namely $\delta
R_{\tau} = \pm 130 \, a^{4}_{\tau}$ \cite{cc} and $\delta R_{\tau} =
(78 \pm 25) a^{4}_{\tau}$ \cite{pp},
and is smaller than
the result of applying the Pad\'e resummation technique directly to
$R_{\tau}$, namely $\delta R_{\tau} = 133 a^{4}_{\tau}$ \cite{sli},
\cite{samn}.

10.~~
 The application of the Pad\'e resummation
technique to $R(s)$  \cite{sli}, stimulated by the previous similar
studies of Ref. \cite{jji}, gives less definite estimates than the
results of applications of our methods (compare the estimate $\delta
R(s) = \left( -49^{+54}_{-40} \right ) a^{4}$ \cite{sli} with the
result $\delta R(s) = -97 a^{4}$ from Table 2 and Eq. (39)). Probably,
this fact
is related with the problems of applicability of the Pad\'e resummation
technique to the sign variation perturbative series for $R(s)$ in the
$\overline{MS}$-scheme.

11.~~
Let us emphasize again, that the used by us method was applyed directly
to the Euclidean quantities, namely to the $D$-function and
deep-inelastic scattering sum rules. The results for $R(s)$ and
$R_{\tau}$ were obtained from the ones of the $D$-function after
taking into account explicitely calculable terms of the analytic
continuation to the Minkowskian region. In principle, one can
try to study the application of the procedure used by us directly to
$R(s)$ and $R_{\tau}$ in the Minkowskian region. We checked that
in the case of applications of Eqs. (6)-(8) with the substitution
of $d_i=r_i$ or $d_i=r_i^{\tau}$
one can reproduce the same values of the Euclidean
coefficients $d_i$ but the structure of the high order $\pi^2$-terms
will not agree with the explicitly known results.
A similar problem will also definitely arise in the case
of more rigorous studies of the applicability of the Pad\'e
resummation methods for the Minkowskian quantities $R(s)$ and
$R_{\tau}$ (for the  studies existing at present see Refs.
\cite{sli,samn}). We think that these
analyses should be supplemented by the construction of the
Pad\'e approximants directly to the $D$-function. In our case the
solution of this problem lies in the necessity of modifications of
Eqs. (6)-(8) in the Minkowskian region by adding concrete calculable
$\pi^2$-dependent and scheme-independent factors.
 After these
modifications it is possible to reproduce our results for
$R(s)$ and $R_{\tau}$ by means of application of
 a similar technique directly in the Minkowskian region.

12.~~
 Our estimate for $d_3(f=3)$ was recently
supported by the phenomenological analysis of the ALEPH data for
$R_{\tau}$ \cite{Dib}. The above presented
considerations of the diffence
between the structure of the perturbative series for the
$D$-funcion and $R(s)$ already stimulated the reconsiderations
of the applications of the Pad\'e approximants for the
estimates of the higher-order coefficients of $R(s)$ and
$R_{\tau}$ \cite{sam}. Note, however, that we are realizing
that at the present level of understanding both methods suffer
from the lack of more rigorous dynamical
information about the analytical
structure of the expanded in the perturbative series functions
both in the Euclidean and Minkowskian regions. We hope that
the considerations presented above will stimulate future more
detailed studies of the related problems.

13.~~~The interesting fact is that the presented $NNLO$ estimates
are in better agreement with the explicite results for $f=3$
numbers of flavours than say for $f=5$ numbers of flavours.
It should be stressed again that the basic assumption of our
approach is the relation $c_N^{ECH}=c_N$. However, it is known
that on the contary to the coefficients $c_N$
the scheme-invariants $c_N^{ECH}$ contain additional factors
of order $f^{N+1}$ \cite{Kato}
, though with small coefficients. For larger
values of $f$ these terms become more and more important. This might
lead to the additional sourse of the violation of the basic
relation $c_N^{ECH}=c_N$ of the re-expansion procedure used by us.

{\bf Acknowledgements}

We are grateful to R.N. Faustov for attracting our attention to the
 the results of Ref. \cite{ss} at the
preliminary stage  of our similar QED studies, which will be presented
elsewhere \cite{fut},  to J. Ch\'yla, F. Le Diberder, J. Ellis,
 G. Marchesini, A. Peterman and M. Karliner for the interest in this
work and useful  discussions.  We also wish to thank G. Altarelli
for some constructive remarks.

The work of one of us (V.V.S.) was supported during his stay in Moscow
by the Russian Fund for Fundamental Research, Grant No. 93-02-14428.

\newpage
\begin{center}
\begin{tabular}{|c|c|c|c|c|c|} \hline
$f$ & $d^{ex}_{2}$ & $(d^{est}_{2})_{ECH}$ & $(d^{est}_{2})_{PMS}$ &
$d^{est}_{3}$ & $ (d_4^{est})_{ECH}-c_3d_1$\\ \hline
1 & 14.11 & 7.54 & 7.70 & 75 & 476 \\ \hline
2 & 10.16 & 6.57 & 7.55 & 50 & 260 \\  \hline
3 & 6.37 & 5.61 & 6.40 & 27.5 & 111 \\ \hline
4 & 2.76 & 4.68 & 5.27 & 8.4  & 22.7 \\  \hline
5 & -0.69 & 3.77 & 4.16 & -7.7 & -13.2  \\ \hline
6 & -3.96 & 2.88 & 3.08 & -21 & -2.76  \\ \hline
\end{tabular}
\end{center}
\vspace*{0.5mm}
Table 1 : The results of estimates of the $NNLO$, $N^3LO$ and $N^4LO$
 corrections in
the series for the
$D$-functions.
\begin{center}
\begin{tabular}{|c|c|c|c|c|c|} \hline
$f$ & $r^{ex}_{2}$ & $(r^{est}_{2})_{ECH}$ & $(r^{est}_{2})_{PMS}$ &
$r^{est}_{3}$ & $(r_4^{est})_{ECH}-c_3d_1$\\ \hline
1 & -7.84 & -14.41 & -14.25 & -166 & -1750\\ \hline
2 & -9.04 & -12.65 & -11.67 & -147 & -1161 \\ \hline
3 & -10.27 & -11.04 & -10.25 & -128 & -668 \\ \hline
4 & -11.52 & -9.59 & -9 & -112 & -263 \\ \hline
5 & -12.76 & -8.32 & -7.93 & -97 & 67 \\ \hline
6 & -14.01 & -7.19 & -6.99 & -83 & 330 \\ \hline
\end{tabular}
\end{center}
\vspace*{0.5mm}
Table 2: The results of estimates of the $NNLO$, $N^3LO$ and $N^4LO$
 corrections in
the series for $R(s)$.
\begin{center}
\begin{tabular}{|c|c|c|c|c|c|} \hline
$f$ & $d^{ex}_{2}$ & $(d^{est}_{2})_{ECH}$ & $(d^{est}_{2})_{PMS}$ &
$d^{est}_{3}$ & $(d_4^{est})_{ECH}-c_3d_1$ \\ \hline
1 & 44.97 & 39.62 & 40.78 & 424 & 4127 \\ \hline
2 & 36.19 & 33.28 & 34.26& 303 & 2613 \\ \hline
3 & 27.89 & 27.37 & 28.16 & 200 & 1474 \\ \hline
4 & 20.07 & 21.91 & 22.50 & 114 & 664 \\ \hline
5 & 12.72 & 16.91 & 17.30 & 44 & 138 \\ \hline
6 & 5.85 & 12.39 & 12.59 & -10 & -145 \\ \hline
\end{tabular}
\end{center}
\vspace*{0.5mm}
Table 3: The results of estimates of the $NNLO$, $N^3LO$ and $N^4LO$
 corrections in
the series for BjnSR.
\begin{center}
\begin{tabular}{|c|c|c|c|c|c|} \hline
$f$ & $d^{ex}_{2}$ & $(d^{est}_{2})_{ECH}$ & $(d^{est}_{2})_{PMS}$ &
$d^{est}_{3}$ & $(d_4^{est})_{ECH}-c_3d_1$\\ \hline
1 & 34.01 & 27.25 & 28.41 & 290 & 2561\\ \hline
2 & 26.93 & 23.11 & 24.09 & 203 & 1580 \\ \hline
3 & 20.21 & 19.22 & 20.01 & 130 & 852\\ \hline
4 & 13.84 & 15.57 & 16.16 & 68 & 343 \\ \hline
5 & 7.83 & 12.19 & 12.59& 18 & 25 \\ \hline
6 & 2.17 & 9.08 & 9.29 & -22 & -130 \\ \hline
\end{tabular}
\end{center}
\vspace*{0.5mm}
Table 4: The results of estimates of the $NNLO$, $N^3LO$ and $N^4LO$
 corrections in
the series for BjpSR and GLSSR.

\end{document}